\documentclass{iau}
\usepackage{graphicx, bm, amssymb, xcolor}
\usepackage{hyperref}
\usepackage{subfigure}

\usepackage{verbatim}
\usepackage[all]{hypcap}
\usepackage{xspace}
\usepackage{multirow}

\usepackage{epsfig}
\usepackage{natbib}

\title[CMZoom Survey]
{A Brief Update on the \textit{CMZoom} Survey}

\author[Battersby et al.]   
{C. Battersby$^1$, E. Keto$^1$, Q. Zhang$^1$, S.N. Longmore$^2$,
J.M.D. Kruijssen$^3$, T. Pillai$^4$, J. Kauffmann$^4$, 
D. Walker$^{1,2}$, X. Lu$^{1,5}$, A. Ginsburg$^6$, J. Bally$^7$,
E.A.C. Mills$^8$, J. Henshaw$^2$, K. Immer$^6$, N. Patel$^1$,
V. Tolls$^1$, A. Walsh$^9$, K. Johnston$^{10}$, \and L.C. Ho$^{11}$}

\affiliation{$^1$Harvard-Smithsonian Center for Astrophysics, U.S.A. 
$^2$ Astrophysics Research Institute, Liverpool John Moores University, U.K. 
$^3$ Astronomisches Rechen-Institut, Zentrum fur Astronomie der Universitat Heidelberg, Germany
$^4$ Max Planck Institut fur Radioastronomie, Auf dem Hugel 69, D-53121 Bonn, Germany
$^5$ School of Astronomy and Space Science, Nanjing University, Nanjing, Jiangsu 210093, China
$^6$ European Southern Observatory, Karl-Schwarzschild-Strasse 2, D-85748 Garching bei München, Germany
$^7$ Centre for Astrophysics and Space Astronomy, University of Colorado, U.S.A.
$^8$ San Jose State University, 1 Washington Square, San Jose, CA 95192, U.S.A.
$^9$ International Centre for Radio Astronomy Research, Curtin University, GPO Box U1987, Perth WA 6845, Australia
$^{10}$ School of Physics \& Astronomy, E.C. Stoner Building, The University of Leeds, Leeds, LS2 9JT, U.K.
$^{11}$ Kavli Institute for Astronomy and Astrophysics, Peking University, Beijing 100871, China
}

\pubyear{2016}
\volume{322}
\pagerange{}

\jname{Astrophysics of the Galactic Centre}
\editors{R.~Crocker, S.~Longmore \& G.~Bicknell}

\begin{document}

\def\Msun{\hbox{M$_{\odot}$}}
\def\kms{km~s$^{\rm -1}$}
\def\micron{$\mu$m}
\def\13CO{$^{13}$CO}
\def\deg{$^{\circ}$}
\def\arcsec{$^{\prime\prime}$}
\def\arcmin{$^{\prime}$}
\def\vlsr{\hbox{V$_{LSR}$}}

\maketitle

\begin{abstract}
The inner few hundred parsecs of the Milky Way, the Central Molecular Zone (CMZ), is our closest laboratory for understanding star formation in the extreme environments (hot, dense, turbulent gas) that once dominated the universe.  We present an update on the first large-area survey to expose the sites of star formation across the CMZ at high-resolution in submillimeter wavelengths: the \textit{CMZoom} survey with the Submillimeter Array (SMA).  We identify the locations of dense cores and search for signatures of embedded star formation.  \textit{CMZoom} is a three-year survey in its final year and is mapping out the highest column density regions of the CMZ in dust continuum and a variety of spectral lines around 1.3 mm.  \textit{CMZoom} combines SMA compact and subcompact configurations with single-dish data from BGPS and the APEX telescope, achieving an angular resolution of about 4\arcsec~(0.2 pc) and good image fidelity up to large spatial scales.  
 
\end{abstract}

\firstsection

\section{Introduction} \label{sec:intro}

The inner few hundred parsecs of the Milky Way, the Central Molecular Zone (CMZ), contains a vast reservoir of dense, molecular gas that is underproducing stars \citep[e.g.][]{imm12,yus09} by about an order of magnitude \citep{lon13a}.
Moreover, this gas is an extreme environment with high gas pressures, densities, and temperatures \citep{mor96, gin16, mil13}, reminiscent of the conditions seen at high redshift \citep{kru13} and ideal for testing prescriptions for star formation at a distance of only 8.4 kpc \citep{rei14}.  

While decades of observations toward our Galaxy's center have revealed its extreme environment, high levels of optical extinction (Av $\sim$ 10-100) have prevented a detailed characterization of the region at shorter wavelengths, where the native resolution is higher.  
Stars form from cool, collapsing clouds of molecular gas which radiate most of their light at long wavelengths (from the submillimeter $\lambda$ $\sim$ 0.5 - 3 mm, to radio $\lambda$ $>$  3mm).  Therefore, the best ``light" with which to see star formation in our Galactic center is through radio/submillimeter observations.  Previous radio and submillimeter observations have been limited to low-resolution \citep[$\sim$ 1 pc; e.g.][]{bal87, jon12} or very limited fields of view \citep[$<$ 10 parsecs$^2$; e.g.][]{rat14}.  

The \textit{CMZoom} survey\footnote{\href{https://www.cfa.harvard.edu/sma/LargeScale/CMZ/}{https://www.cfa.harvard.edu/sma/LargeScale/CMZ/}} fills this gap by observing the CMZ at submillimeter wavelengths over a large area with high spatial resolution.  \textit{CMZoom} allows us to identify the
present and future sites of star formation (through their cold, dust continuum emission) and characterize the physical properties (turbulent line widths and gas temperatures using H$_2$CO) and star-forming activity (e.g. CH$_3$CN as a tracer of hot core chemistry and SiO as a tracer of shocks in bipolar outflows) of the protostars found there.  The survey is now complete and we expect initial papers to be submitted by Spring 2017 and a full data release in late Summer or early Fall 2017.  This survey takes advantage of the unique fast-mapping capabilities of the SMA (due to its large field-of-view) and opens up important follow-up opportunities with the Atacama Large Millimeter Array (ALMA).

\section{Observations}

The \textit{CMZoom} survey on the Submillimeter Array (SMA) was designed to survey a large area of the CMZ, selected for high column density, to uncover and characterize the dense substructure of CMZ gas clouds and their basic physical properties.  The large primary beam (about 50\arcsec) and wide bandwidth ($\Delta$$\nu$ $>$ 8 GHz) of the SMA makes it ideal for a moderately high-resolution (4\arcsec), moderately sensitive (RMS $\sim$ 4 mJy) survey over a very large area of the CMZ, and opens up important follow-up opportunities with complementary facilities.

\begin{figure}[b]
\begin{center}
\subfigure{ 
\includegraphics[trim=0 30mm 0 0, clip, width=1\textwidth]{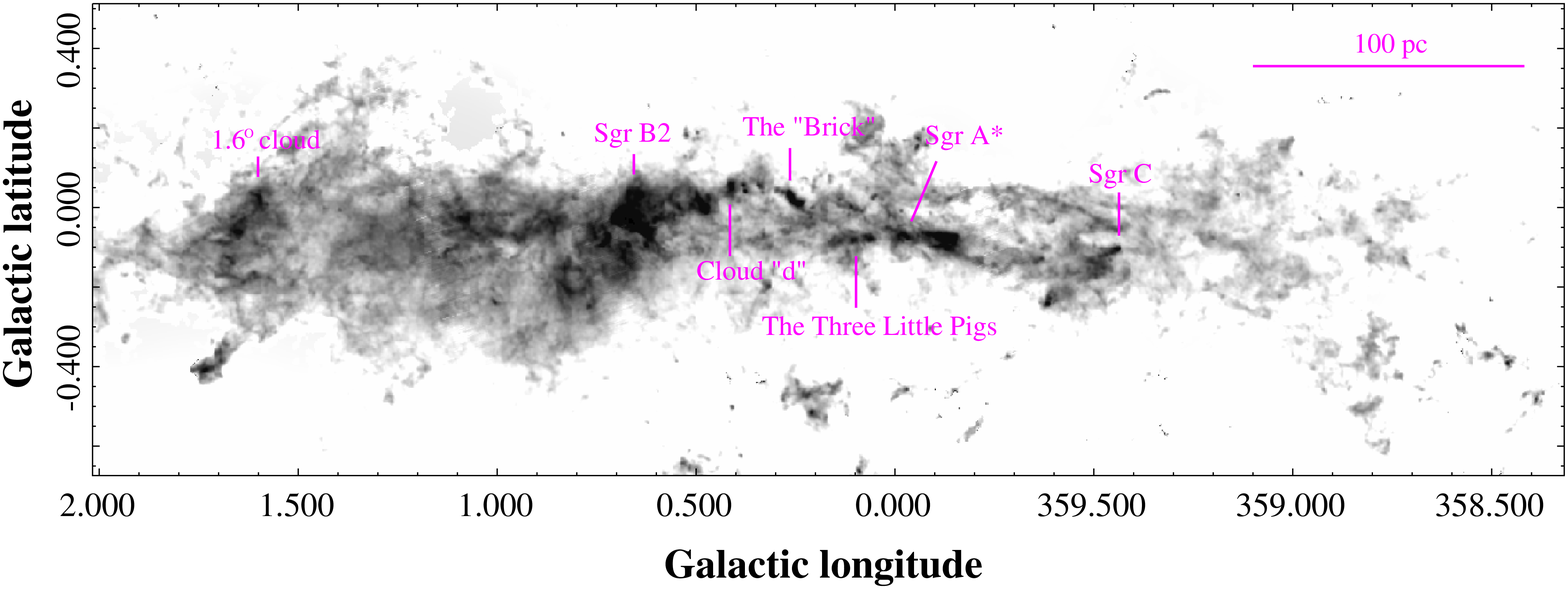}}\\
\vspace{-5mm}
\subfigure{
\includegraphics[trim=0 0 0 4mm, clip, width=1\textwidth]{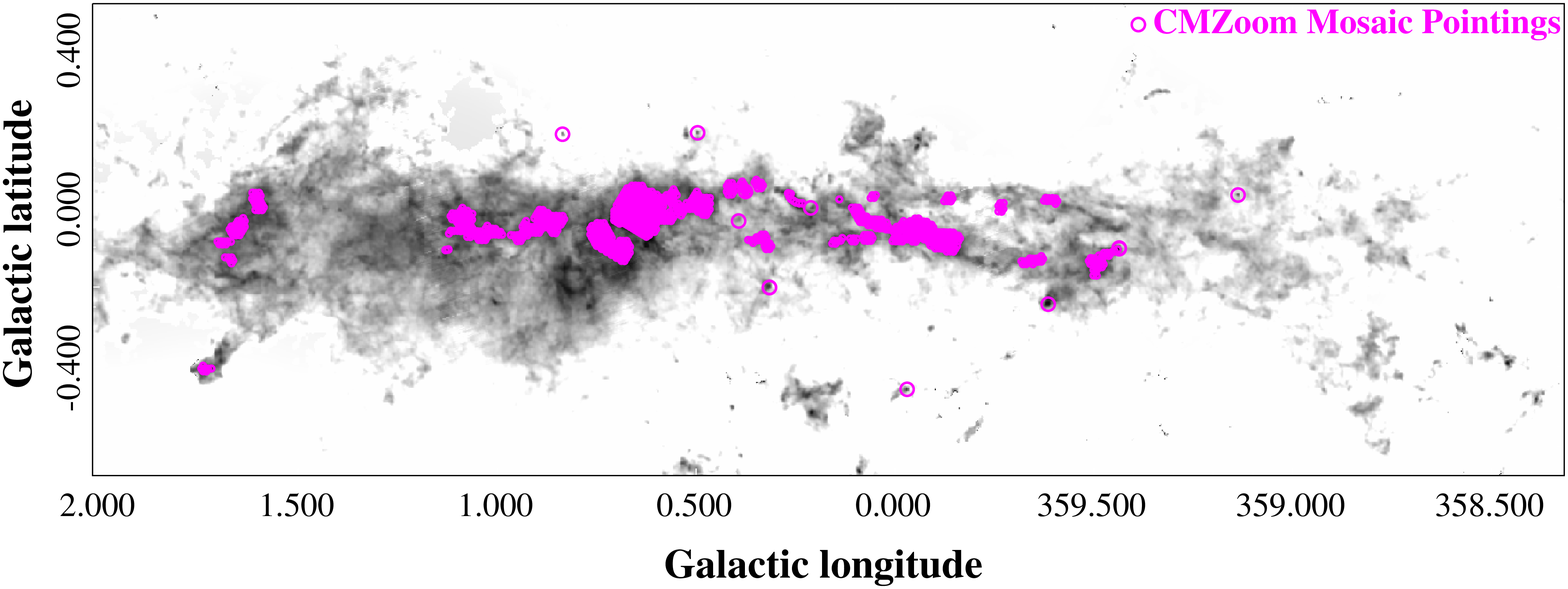}}\\
 \caption{The Central Molecular Zone in N(H$_2$) \citep[Battersby et al., in prep.][]{mol11} with select regions labelled in the top figure.  The bottom figure shows the specific mosaic pointings for \textit{CMZoom}.}
\label{fig1}
\end{center}
\end{figure}

\subsection{Coverage}

The \textit{CMZoom} target area was selected by H$_2$ column density maps derived from Herschel emission \citep[Battersby et al., in prep., based on the algorithms and data from][]{bat11,mol10}.  \textit{CMZoom} covers the entire CMZ above an H$_2$ column density of 10$^{23}$ cm$^{-2}$, with a few exceptions for known foreground regions and the extended, negative latitude envelope near Sgr B2 (see Figure 1).  The \textit{CMZoom} target area also includes a number of pointings toward CMZ candidate regions of `isolated high-mass star formation,' a few selected regions with N(H$_2$) $>$ 3$\times$10$^{22}$ cm$^{-2}$ in the inner 100 pc streams, the circumnuclear disk (CND) and its connection to the 20 and 50 \kms~clouds, and a bridge of molecular gas detected in H$_2$CO from \citet{gin16} near [$\ell$,b]=[0.06\deg,-0.04\deg].

\subsection{Submillimeter Array}

SMA observations of the target area were completed in 61 tracks from the period of May 20, 2014 to September 24, 2016.  Each night, about 20 pointings of the large mosaic (see Figure 1) were observed, repeating throughout the night for good UV coverage.  Weather conditions were typically better than 2.5 mm of precipitable water vapor, but there was some variation, and sometimes multiple poorer weather or shorter tracks would be combined to achieve better overall RMS.  Due to the varying weather conditions and timing of observations (during some periods the Galactic Center is visible for longer periods), the RMS of the survey is not uniform, but is generally between about 3-5 mJy RMS in the 1.3 mm dust continuum.  45 compact tracks and 16 subcompact tracks were observed.  These observations were calibrated using MIR IDL with scripts designed for the SMA.  The compact, subcompact, and single-dish data were cleaned and combined using CASA.  The final \textit{CMZoom} resolution is about 4\arcsec \citep[or 0.2 pc at a distance of 8.4 kpc][]{rei14}.

\textit{CMZoom} took advantage of both the ASIC and SWARM correlators, as the SMA transitioned between them over the period of the survey.  Our primary setup in the original ASIC correlator had two sidebands, the lower sideband covering from 216.9 GHz to 220.9 GHz and the upper sideband covering the frequency range of 228.9 GHz to 232.9 GHz.  The lower sideband coverage includes SiO (5-4), $^{13}$CO and C$^{18}$O (2-1), three H$_2$CO transitions near 218 GHz, a number of CH$_3$OH lines, and more.  The upper sideband coverage includes the CO 2-1 transition, H$\alpha$30, and a number of CH$_3$OH transitions.  Our velocity resolution is 1.1 \kms.  The SWARM correlator coverage varied over the course of the survey, from 0 GHz additional coverage to 8 GHz of added frequency coverage at high spectral resolution.

\subsection{Single Dish}

We combine our SMA compact and subcompact observations with single-dish data to achieve better recovery of structure at large spatial scales.
For the dust continuum emission, we use data from the Bolocam Galactic Plane Survey \citep[BGPS;][]{agu11,gin13} -- single-dish observations on the Caltech Submillimeter Observatory of the 1.1mm dust continuum.  APEX data from \citet{gin16} complement our spectral line data in the lower sideband from about 216-220 GHz.  Most of the \textit{CMZoom} diagnostically important transitions (e.g. H$_2$CO, $^{13}$CO, SiO) are in the lower sideband, and are combined with this single-dish data.  The upper sideband spectral lines (namely CO 2-1) are not complemented by single-dish data, but are found to be too extended to be of very much use for deriving physical or kinematic properties of the clouds.

\begin{figure}[h]
\includegraphics[trim=0 5mm 0 0mm, clip, width=1\textwidth]{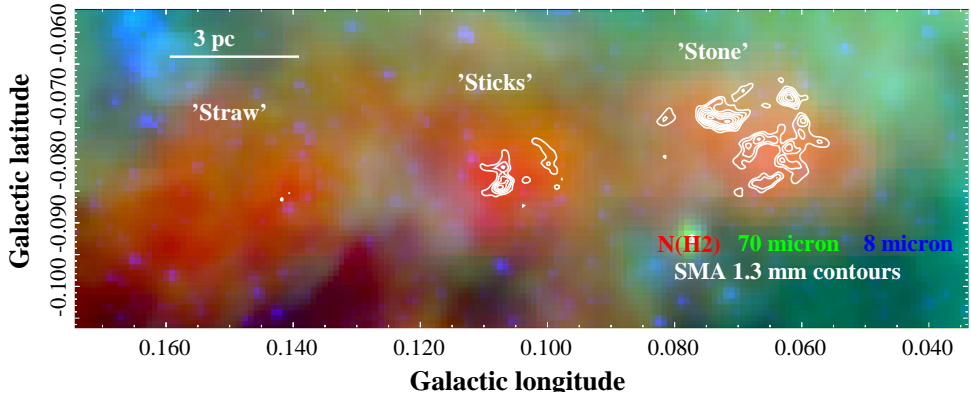}
\caption{\em{\small{While extremely similar in their global properties, 
observations with the SMA reveal remarkably different substructure in the Three Little Pigs clouds.  From widespread non-detections in the `straw' cloud, to moderate substructure in the `sticks' cloud and a complicated network of filaments and cores in the `Stone' cloud, the origin of this variety is yet unknown.  
The Three Little Pigs are shown here in a three-color image (Red: Herschel PACS/SPIRE cold dust column density, Green: 70 \micron, Blue: GLIMPSE 8 \micron) while contours from the \textit{CMZoom} survey are shown in white, the lowest contour at 5$\sigma$.}}}
\end{figure}

\section{Survey Goals \& Preliminary Highlights}

The \textit{CMZoom} survey was designed to: 1) identify the location of dense, embedded cores in the CMZ, 2) determine their physical and kinematic properties, and 3) search for star-forming activity toward these cores.  Observations of the dust continuum and dense gas tracer H$_2$CO help to accomplish goal 1, while temperature determinations with H$_2$CO, as well as line widths and kinematic motions, observations of SiO 5-4, and the dust continuum help to accomplish goal 2.  Goal 3 is accomplished with a variety of tracers, including searching for outflows in SiO and $^{13}$CO, compact hot core emission in CH$_3$OH, looking for high-density power-law tails in N-PDFs of the dust continuum emission, and low virial ratios using H$_2$CO line wdiths.

The scientific goals of the survey are to: 1) Identify the present and future sites of star formation, 2) Answer why the star formation rate is so low in this region, and if we can better `prescribe' star formation universally with consideration for environment, 3) Answer if star formation can be triggered by pericenter passage by the bottom of the global gravitational potential well, 
and 4) Find the precursors to the most massive stars in the Galaxy.  Figure 2 shows a brief highlight of SMA dust continuum observations toward the ``Three Little Pigs" clouds\footnote{The names for these three clouds were inspired by a fable, in each of three little pigs builds a house to protect itself from the Big Bad Wolf; one of straw, one of sticks, and one of stone.} and \citet{lu15} highlight early results toward the 20 \kms~cloud.

\section*{Acknowledgements}
We thank the organizers for a productive and stimulating meeting.  We thank the staff of the SMA -- the operators, software developers, the director and time allocation managers, schedulers, scientists, and those upgrading the SMA instrumentation for their endless efforts and support, without whom, this project would not have been possible.  This work has made use of MIR IDL, CASA, glue, ds9, and astropy.  C.B. is supported by the National Science Foundation under Award No. 1602583.  

\newpage
\bibliographystyle{iau}
\bibliography{references1}

\begin{thebibliography}{}
\makeatletter
\relax
\def\mn@urlcharsother{\let\do\@makeother \do\$\do\&\do\#\do\^\do\_\do\%\do\~}
\def\mn@doi{\begingroup\mn@urlcharsother \@ifnextchar [ {\mn@doi@}
  {\mn@doi@[]}}
\def\mn@doi@[#1]#2{\def\@tempa{#1}\ifx\@tempa\@empty \href
  {http://dx.doi.org/#2} {doi:#2}\else \href {http://dx.doi.org/#2} {#1}\fi
  \endgroup}
\def\mn@eprint#1#2{\mn@eprint@#1:#2::\@nil}
\def\mn@eprint@arXiv#1{\href {http://arxiv.org/abs/#1} {{\tt arXiv:#1}}}
\def\mn@eprint@dblp#1{\href {http://dblp.uni-trier.de/rec/bibtex/#1.xml}
  {dblp:#1}}
\def\mn@eprint@#1:#2:#3:#4\@nil{\def\@tempa {#1}\def\@tempb {#2}\def\@tempc
  {#3}\ifx \@tempc \@empty \let \@tempc \@tempb \let \@tempb \@tempa \fi \ifx
  \@tempb \@empty \def\@tempb {arXiv}\fi \@ifundefined
  {mn@eprint@\@tempb}{\@tempb:\@tempc}{\expandafter \expandafter \csname
  mn@eprint@\@tempb\endcsname \expandafter{\@tempc}}}

\bibitem[\protect\citeauthoryear{{Aguirre} et~al.,}{{Aguirre}
  et~al.}{2011}]{agu11}
{Aguirre} J.~E.,  et~al., 2011, \mn@doi [\apjs] {10.1088/0067-0049/192/1/4},
  \href {http://adsabs.harvard.edu/abs/2011ApJS..192....4A} {192, 4}

\bibitem[\protect\citeauthoryear{{Bally} et~al.,}{{Bally} et~al.}{1987}]{bal87}
{Bally} J.,  et~al., 1987, \mn@doi [\apjs] {10.1086/191217}, \href
  {http://adsabs.harvard.edu/abs/1987ApJS...65...13B} {65, 13}

\bibitem[\protect\citeauthoryear{{Battersby} et~al.,}{{Battersby}
  et~al.}{2011}]{bat11}
{Battersby} C.,  et~al., 2011, \mn@doi [\aap] {10.1051/0004-6361/201116559},
  \href {http://adsabs.harvard.edu/abs/2011A%26A...535A.128B} {535, A128}

\bibitem[\protect\citeauthoryear{{Ginsburg} et~al.,}{{Ginsburg}
  et~al.}{2013}]{gin13}
{Ginsburg} A.,  et~al., 2013, \mn@doi [\apjs] {10.1088/0067-0049/208/2/14},
  \href {http://adsabs.harvard.edu/abs/2013ApJS..208...14G} {208, 14}

\bibitem[\protect\citeauthoryear{{Ginsburg} et~al.,}{{Ginsburg}
  et~al.}{2016}]{gin16}
{Ginsburg} A.,  et~al., 2016, \mn@doi [\aap] {10.1051/0004-6361/201526100},
  \href {http://adsabs.harvard.edu/abs/2016A%26A...586A..50G} {586, A50}

\bibitem[\protect\citeauthoryear{{Immer} et~al.,}{{Immer} et~al.}{2012}]{imm12}
{Immer} K.,  et~al., 2012, \mn@doi [\aap] {10.1051/0004-6361/201117857}, \href
  {http://adsabs.harvard.edu/abs/2012A%26A...537A.121I} {537, A121}

\bibitem[\protect\citeauthoryear{{Jones} et~al.,}{{Jones} et~al.}{2012}]{jon12}
{Jones} P.~A.,  et~al., 2012, \mn@doi [\mnras]
  {10.1111/j.1365-2966.2011.19941.x}, \href
  {http://adsabs.harvard.edu/abs/2012MNRAS.419.2961J} {419, 2961}

\bibitem[\protect\citeauthoryear{{Kruijssen} \& {Longmore}}{{Kruijssen} \&
  {Longmore}}{2013}]{kru13}
{Kruijssen} J.~M.~D.,  {Longmore} S.~N.,  2013, \mn@doi [\mnras]
  {10.1093/mnras/stt1634}, \href
  {http://adsabs.harvard.edu/abs/2013MNRAS.435.2598K} {435, 2598}

\bibitem[\protect\citeauthoryear{{Longmore} et~al.,}{{Longmore}
  et~al.}{2013}]{lon13a}
{Longmore} S.~N.,  et~al., 2013, \mn@doi [\mnras] {10.1093/mnras/sts376}, \href
  {http://adsabs.harvard.edu/abs/2013MNRAS.429..987L} {429, 987}

\bibitem[\protect\citeauthoryear{{Lu} et~al.,}{{Lu} et~al.}{2015}]{lu15}
{Lu} X.,  et~al., 2015, \mn@doi [\apjl] {10.1088/2041-8205/814/2/L18}, \href
  {http://adsabs.harvard.edu/abs/2015ApJ...814L..18L} {814, L18}

\bibitem[\protect\citeauthoryear{{Mills} \& {Morris}}{{Mills} \&
  {Morris}}{2013}]{mil13}
{Mills} E.~A.~C.,  {Morris} M.~R.,  2013, \mn@doi [\apj]
  {10.1088/0004-637X/772/2/105}, \href
  {http://adsabs.harvard.edu/abs/2013ApJ...772..105M} {772, 105}

\bibitem[\protect\citeauthoryear{{Molinari} et~al.,}{{Molinari}
  et~al.}{2010}]{mol10}
{Molinari} S.,  et~al., 2010, \mn@doi [\aap] {10.1051/0004-6361/201014659},
  \href {http://adsabs.harvard.edu/abs/2010A%26A...518L.100M} {518, L100}

\bibitem[\protect\citeauthoryear{{Molinari} et~al.,}{{Molinari}
  et~al.}{2011}]{mol11}
{Molinari} S.,  et~al., 2011, \mn@doi [\apjl] {10.1088/2041-8205/735/2/L33},
  \href {http://adsabs.harvard.edu/abs/2011ApJ...735L..33M} {735, L33}

\bibitem[\protect\citeauthoryear{{Morris} \& {Serabyn}}{{Morris} \&
  {Serabyn}}{1996}]{mor96}
{Morris} M.,  {Serabyn} E.,  1996, \mn@doi [\araa]
  {10.1146/annurev.astro.34.1.645}, \href
  {http://adsabs.harvard.edu/abs/1996ARA%26A..34..645M} {34, 645}

\bibitem[\protect\citeauthoryear{{Rathborne} et~al.,}{{Rathborne}
  et~al.}{2014}]{rat14}
{Rathborne} J.~M.,  et~al., 2014, \mn@doi [\apjl]
  {10.1088/2041-8205/795/2/L25}, \href
  {http://adsabs.harvard.edu/abs/2014ApJ...795L..25R} {795, L25}

\bibitem[\protect\citeauthoryear{{Reid} et~al.,}{{Reid} et~al.}{2014}]{rei14}
{Reid} M.~J.,  et~al., 2014, \mn@doi [\apj] {10.1088/0004-637X/783/2/130},
  \href {http://adsabs.harvard.edu/abs/2014ApJ...783..130R} {783, 130}

\bibitem[\protect\citeauthoryear{{Yusef-Zadeh} et~al.,}{{Yusef-Zadeh}
  et~al.}{2009}]{yus09}
{Yusef-Zadeh} F.,  et~al., 2009, \mn@doi [\apj] {10.1088/0004-637X/702/1/178},
  \href {http://adsabs.harvard.edu/abs/2009ApJ...702..178Y} {702, 178}

\makeatother
\end{thebibliography}

%
%

\end{document}